\documentclass[twocolumn,tighten]{aastex701}
\usepackage{multirow}
\usepackage{microtype,booktabs}
\usepackage{bm}

\graphicspath{ {./figures/}{./} }

\begin{document}

\title{A JWST Paschen-$\alpha$ Calibration of the Radio Luminosity-Star Formation Rate Relation at $z\sim 1.3$}

\author[0000-0003-3506-5536]{Nick Seymour}
\email{nick.seymour@curtin.edu.au}
\affiliation{International Centre for Radio Astronomy Research, Curtin University, GPO Box U1987, Bentley, WA 6845, Australia}

\author[0000-0001-6279-4772]{Catherine Hale}
\email{catherine.hale@physics.ox.ac.uk}
\affiliation{Astrophysics, Department of Physics, University of Oxford, Keble Road, Oxford, OX1 3RH, UK}
\affiliation{Institute for Astronomy, University of Edinburgh,  Royal Observatory Edinburgh, Blackford Hill, Edinburgh, EH9 3HJ, UK}

\author[0000-0003-2265-5983]{Imogen Whittam}
\email{imogen.whittam@physics.ox.ac.uk}
\affiliation{Astrophysics, Department of Physics, University of Oxford, Keble Road, Oxford, OX1 3RH, UK}
\affiliation{Department of Physics \& Astronomy, University of the Western Cape, Private Bag X17, Bellville, Cape Town, 7535, South Africa}

\author[0000-0001-5851-6649]{Pascal Oesch}
\email{pascal.oesch@unige.ch}
\affiliation{Department of Astronomy, University of Geneva, Chemin Pegasi 51, CH-1290 Versoix, Switzerland}
\affiliation{Cosmic Dawn Center (DAWN), Niels Bohr Institute, University of Copenhagen, Jagtvej 128, DK-2200 K{\o}benhavn N, Denmark}

\author[0000-0002-9672-3005]{Alba Covelo-Paz}
\email{Alba.Covelopaz@unige.ch}
\affiliation{Department of Astronomy, University of Geneva, Chemin Pegasi 51, 1290 Versoix, Switzerland}

\author[0000-0003-3735-1931]{Stijn Wuyts}
\email{sw2122@bath.ac.uk}
\affiliation{Department of Physics, University of Bath, Claverton Down, Bath, BA2 7AY, UK}
\author[0000-0002-9149-2973]{J. Afonso}
\email{jmafonso@ciencias.ulisboa.pt}
\affiliation{Instituto de Astrofísíca e Ciências do Espaço, Faculdade de Ciências, Universidade de Lisboa, OAL, Tapada da Ajuda, PT1349-018 Lisboa, Portugal}

\author[0000-0003-3917-1678]{R. A. A. Bowler}
\email{rebecca.bowler@manchester.ac.uk}
\affiliation{Jodrell Bank Centre for Astrophysics, Department of Physics and Astronomy, School of Natural Sciences, The University of Manchester, Manchester, M13 9PL, UK}

\author[0000-0002-4440-8046]{Joe Arthur Grundy}
\email{joe.grundy@postgrad.curtin.edu.au}
\affiliation{International Centre for Radio Astronomy Research, Curtin University, GPO Box U1987, Bentley, WA 6845, Australia}

\author[0000-0003-2035-3850]{Ravi Jaiswar}
\email{ravi.jaiswar@postgrad.curtin.edu.au}
\affiliation{International Centre for Radio Astronomy Research, Curtin University, GPO Box U1987, Bentley, WA 6845, Australia}

\author[0000-0001-7039-9078]{Matt Jarvis}
\email{matt.jarvis@physics.ox.ac.uk}
\affiliation{Astrophysics, Department of Physics, University of Oxford, Keble Road, Oxford, OX1 3RH, UK}
\affiliation{Department of Physics \& Astronomy, University of the Western Cape, Private Bag X17, Bellville, Cape Town, 7535, South Africa}

\author[0000-0002-6479-6242]{Allison Matthews}
\email{amatthews@carnegiescience.edu}
\affiliation{Carnegie Observatories, 813 Santa Barbara Street, Pasadena, CA 91101, USA}

\author[0000-0001-5492-4522]{Romain A. Meyer}
\email{Romain.Meyer@unige.ch}
\affiliation{Department of Astronomy, University of Geneva, Chemin Pegasi 51, 1290 Versoix, Switzerland}

\author[0000-0002-6558-9894]{Chloe Neufeld}
\email{chloe.neufeld@yale.edu}
\affiliation{Astronomy Department, Yale University, 52 Hillhouse Avenue, New Haven, CT 06511, USA}

\author[0000-0001-9687-4973]{Naveen A. Reddy}
\email{naveenr@ucr.edu}
\affiliation{Department of Physics \& Astronomy, University of California, Riverside, CA 92521, USA}

\author[0000-0003-4702-7561]{Irene Shivaei}
\email{irene.shivaei@gmail.com}
\affiliation{Centro de Astrobiolog\'{i}a (CAB), CSIC-INTA, Carretera de Ajalvir km 4, Torrej\'{o}n de Ardoz, 28850, Madrid, Spain}

\author[0000-0001-9708-253X]{Dan Smith}
\email{daniel.j.b.smith@gmail.com}
\affiliation{Centre for Astrophysics Research, Department of Physics, Astronomy and Mathematics, University of Hertfordshire, Hatfield AL10 9AB, UK}

\author[0009-0006-9953-6471]{Rohan Varadaraj}
\email{rohan.varadaraj@physics.ox.ac.uk}
\affiliation{Astrophysics, Department of Physics, University of Oxford, Keble Road, Oxford, OX1 3RH, UK}

\author[0000-0002-1033-3656]{Michael A. Wozniak}
\email{michael.wozniak@email.ucr.edu}
\affiliation{Department of Physics and Astronomy, University of California, Riverside, 900 University Avenue, Riverside, CA 92521, USA}

\author[0000-0001-5512-3735]{Lyla Jung}
\email{lyla.jung@physics.ox.ac.uk}
\affiliation{Astrophysics, Department of Physics, University of Oxford, Keble Road, Oxford, OX1 3RH, UK}



\begin{abstract} 
As radio emission from normal galaxies is a dust-free tracer of star formation, tracing the star formation history of the Universe is a key goal of the SKA and ngVLA. In order to investigate how well radio luminosity traces star formation rate (SFR) in the early Universe, we have examined the radio properties of a {\it JWST} Paschen-$\alpha$ sample of galaxies at $1.0\lesssim z \lesssim 1.8$. In the GOODS-S field, we cross-matched a sample of 506 FRESCO Paschen-$\alpha$ emitters with the 1.23\,GHz radio continuum data from the MeerKAT MIGHTEE survey finding 47 detections. After filtering for AGN (via X-ray detections,  hot mid-infrared dust and extended radio  emission), as well as blended sources, we obtained a sample of SFGs comprising:  11 cataloged  radio detections,  18 non-cataloged detections (at $\approx3-5\sigma$) and  298 undetected sources. Stacking the  298 undetected sources we obtain a $3.3\sigma$ detection in the radio. This sample, along with a local sample of Paschen-$\alpha$ emitters, lies along previous radio luminosity/SFR relations from local ($<0.2$) to high redshift ($z\sim 1$). Fitting the FRESCO data at $1.0\lesssim z \lesssim 1.8$ we find $\log(L_{\rm 1.4GHz})=(1.31\pm 0.17)\times\log(\rm SFR_{Pa\alpha})+(21.36\pm 0.17)$ which is consistent with other literature relations. We can explain some of the observed scatter in the $L_{\rm 1.4GHz}$/SFR$_{\rm Pa\alpha}$ correlation by a toy model in which the synchrotron emission is a delayed/averaged tracer of the instantaneous Paschen-$\alpha$ SFR by $\sim 10/75\,$Myr. 
\end{abstract}




\section{Introduction} 
\label{sec:intro}

Radio emission from normal galaxies, i.e. those without active galactic nuclei (AGN), has long been heralded as an unbiased tracer of star formation, one not effected by gas or dust. Emission at radio frequencies is dominated by synchroton processes from relativistic electrons and free-free processes from thermal electrons \citep[see review by][]{Condon:92}. The relativistic electrons are thought to be accelerated by shocks from supernovae, which only come from massive OB stars \citep[$>8\,$M$_\odot$,][]{kennicutt:12}, and then gyrate in the galactic magnetic field  whereas the thermal electrons are heated by the same bright stars in HII regions. This framework explains the well-established radio/far-infrared (far-IR) luminosity correlation \citep{Harwit:75,condon:91a,Yun:01} observed for star forming galaxies (SFGs) in the local Universe. The far-IR bump is also directly connected to star formation due to interstellar dust thermally re-radiating absorbed UV/optical emission from star forming regions. Hence, the radio luminosity to star formation rate (SFR) calibration was first derived from the radio/far-IR correlation \cite[e.g.][]{Kennicutt:98, Bell:03}.

Deep radio surveys probing below $\sim 1\,$mJy have long shown an upturn in the source counts at faint flux densities \citep[e.g.][]{Windhorst:84a,Seymour:04} which has been shown to be due to a population of distant SFGs. While robustly distinguishing between radio emission from a SFG or AGN remains an issue \citep{algera:20}, deep radio surveys are now routinely used to measure the star formation history of the Universe \citep[{SFH},][]{Seymour:08,Novak:17,cochrane:23}. Furthermore, the radio/far-IR correlation has been shown to extend to high redshifts \citep[$z>1$,][]{Ivison:10} demonstrating that radio luminosity can trace SFR at high-$z$ well, albeit with some potential evolution \citep{magnelli:15, Delhaize:17, Bonato:21,Delvecchio:21}. Some work has been performed using dust-corrected H$\alpha$ emission lines at $1.4<z<2.6$ \citep{duncan:20} which found the SFR/radio luminosity relation at this epoch was consistent with that found locally. 
The measurement of the radio-derived SFH has been  determined from broad and deep source counts as well as confusion P(D) analysis probing source counts below the detection limit \citep[][]{matthews:21} including radio emission potentially emitted up to $z\sim 6$. The culmination of this work was the adoption by the Square Kilometre Array (SKA) and the Next-Generation Very Large Array (ngVLA) of the key science goal of `probing galaxy evolution in the radio continuum' \citep{prandoni:15,jarvis:15,murphy:19}.

Despite this promise there remain some complications in straight-forwardly using radio luminosities as a tracer of SFR. First, the broad radio spectrum ($0.05-50\,$GHz) is not always a simple power law as different frequency regimes sample different physical processes. While the thermal free-free emission dominates at high frequencies with a spectral index of $\alpha\sim -0.1$ (where the flux density, $S_\nu\propto\nu^\alpha$, across frequency, $\nu$), the non-thermal synchrotron emission dominates at lower frequencies with $\alpha\sim -0.8$, around 1\,GHz. In addition to the spectral flattening at higher frequencies as the thermal emission begins to dominate, there has long been evidence for low-frequency turn-overs \citep{Hummel:91, Clemens:10, Galvin:18, grundy:25}, due to free-free absorption, and also evidence for spectral breaks in the synchrotron emission at $1-10\,$GHz \citep{Clemens:10, Galvin:18, Dey:22, Dey:24}. These variations on the radio spectra are on top of varying thermal fractions \citep[$\sim 1-15\%$,][]{Galvin:18,Dey:22}. 

Second, there is the issue of timescales, with the free-free emission being directly related to the number of ionizing photons (with little dependence on temperature) making it a very immediate tracer of the current SFR. Synchrotron emission is likely a delayed tracer of SFR for two reasons: (1) it takes $10-30$\,Myr for the massive stars to go supernova (thereby producing the relativistic electrons which have been shock-accelerated by Type II supernova), and (2) the typical lifetime of synchrotron emitting electrons in a galaxy with a magnetic field of $B=5\,\mu$G is around 100\,Myrs \citep{Condon:92} although the field in star forming regions and nuclei can be higher \citep[$B=10-16\,\mu$G and $B=30\,\mu$G, respectively,][]{murphy:11}. Third, there is issue of inverse Compton (iC) processes depressing the radio emission of SFGs at higher redshifts \citep[e.g.][found a decrease in the radio emission of $3<z<5$ Lyman-break galaxies consistent with iC losses]{whittam:25}. Hence, in the era of the SKA and ngVLA, there is a clear need to ensure that calibration of the radio luminosity to SFR conversion is robust. 

The Paschen-$\alpha$ line is known as a superb tracer of SFR \cite[e.g.][]{alonsoherrero:06} like other optical and UV recombination lines. While it is weaker than the Balmer H$\alpha$ line at $0.6563\,\mu$m, which is widely adopted as a SFR tracer, it has the advantage that its near-IR wavelength of $1.8756\,\mu$m means it suffers a lot less from dust attenuation. However, this rest-frame wavelength lies between $J$ and $H$-band where water vapour decreases the atmospheric transmittance making ground-based observations very difficult \citep{tateuchi:15}. Hence, the limited studies with Paschen-$\alpha$ have either targetted redshifted galaxies \citep[e.g.][]{hill:96,falcke:98,kim:10}, redshifted ULIRGS \citep[e.g.][]{murphy:99}, or used space-based observations. The Near Infrared Camera and Multi-object Spectrometer (NICMOS) on the {\it Hubble Space Telescope} ({\it HST})  undertook several Pa$\alpha$ studies in the local Universe \citep[e.g.][]{alonsoherrero:06,liu:13}, but due to the high angular resolution of  {\it HST} these observations were insensitive to diffuse Paschen-$\alpha$ emission. Hence, such observations could not reliably obtain total Paschen-$\alpha$ fluxes in the local Universe.

SKA precursors (e.g. MeerKAT) and pathfinders (e.g. eVLA, eMERLIN) are now producing surveys around 1\,GHz with RMS sensitivities $\lesssim 1\,\mu$Jy producing catalogs likely to be dominated by high-$z$ SFGs \cite{vandervlugt:21,Hale:25}. At the same time {\it JWST} has the sensitivity and wavelength coverage to detect Paschen-$\alpha$ at $z>1$ \citep[e.g.][]{Neufeld:24}. Indeed, several studies have looked at the combination of Paschen and Balmer lines to study dust attenuation in {\it JWST}-selected galaxies \citep{reddy:23a,liu:24,lorenz:25,reddy:25} in the early Universe. In this work, we investigate the correlation between radio and Paschen-$\alpha$ luminosity in order to verify current radio luminosity/SFR calibrators.  In \S\ref{sec:data} we present the data used and analyse it \S\ref{sec:analysis}. Our results are presented and discussed in \S\ref{sec:result} and conclude the paper in \S\ref{sec:conclusion}. We assume a flat $\Lambda$ cold dark matter ($\Lambda$CDM) cosmology with $\Omega_{\rm M}=0.3$, $\Omega_\Lambda=0.7$, and $H_0=70\;{\rm km\,s}^{-1}\,{\rm Mpc}^{-1}$. 

\section{Data} \label{sec:data}

\subsection{FRESCO Paschen-\texorpdfstring{$\alpha$}~~catalog}
\label{sec:data:fresco}

The First Reionisation Epoch Spectroscopic Complete Observations survey \citep[FRESCO, Cycle 1 GO-1895,][]{oesch:23} is a cycle 1 {\it JWST} program targeting $\sim 60\,$arcmin$^2$ in each GOODS field with two $2\times 4$ NIRCam/grism mosaics with the F444W filter covering 3.8-5.0 $\mu$m and reaching 5$\sigma$ line flux limits of $\sim2\times10^{-18}\,\rm{erg}\,\rm{s}^{-1}\,\rm{cm}^{-2}$. The slitless/grism data were processed by {\tt grizli}\footnote{\url{https://github.com/gbrammer/grizli/}} in the same way as presented in \cite{Neufeld:24} and \cite{Covelopaz:25}. Sources were pre-selected based on their photometric redshifts, before their continuum-subtracted spectra were extracted and redshifts identified with {\tt grizli}. These identifications visually inspected for possible contamination or misclassification. 
Full details on the FRESCO emission line catalogs will be presented in Covelo-Paz et al., in prep.

Here we use the GOODS-South catalog of Paschen-$\alpha$ line detections with 506 sources across $1.0\lesssim z \lesssim 1.8$ ($\langle z\rangle \sim 1.3$) selected to be $>4\sigma$ and described in \cite{Neufeld:24}. Two of these sources have broad lines with FWHM $>1,000\,$km\,s$^{-1}$ \citep[][]{Sun:25}. Modelling of the spectral energy distribution (SED) of the full sample over $0.4-8.0\,\mu$m with {\tt Bagpipes} \citep{carnall:18} provides the usual galaxy properties of visual attenuation $A_{\rm V}$ \citep[using the][reddening curve]{calzetti:00}, stellar mass, SED-fitted SFR and mass-weighted stellar age. {\tt BAGPIPES} uses the \cite{Bruzual:03} stellar population models and has a variable metallicity (with best fits for this sample generally around half of solar). We use the $A_{\rm V}$ to correct the mildly dust obscured Paschen-$\alpha$ flux densities.  To do so we assume that the nebular attenuation, $A_V^{\rm neb}=A_V^{\rm cont}/0.44$, where $A_V^{\rm cont}$ is the attenuation of the stellar continuum \citep{calzetti:97}, a result confirmed by \cite{Arnaudova:25}. While there is some scatter in the ratio between nebular and stellar attenuation \citep[e.g.,][]{koyama:18} choosing extreme values only changes the correction factor below by $<0.1\,$dex. We also assume that the (nebular) Paschen-$\alpha$ attenuation is $A_{Pa\alpha}^{\rm neb}=A_V^{\rm neb}/6.97$ using the \cite{calzetti:00} reddening curve with $R_V=4.05$. Again, varying the choice of reddening curves only changes the correction factor by up to $0.1-0.2\,$dex.
Hence, the final multiplicative correction factor to the Paschen-$\alpha$ flux densities and luminosities is $10^{0.4\,A_{V}^{\rm cont}/(0.44\times 6.97)}$. The mean value of this correction factor was $1.18$, very similar to that estimated by \cite{Neufeld:24} via a different method\footnote{We search for H-$\alpha$ observations of our sample, but only one Paschen-$\alpha$ source 
\citep[in the GOODS-N field,][]{Skelton:14}
has such data. The nebula attenuation derived from the line ratio is just consistent with that from the BAGPIPES (which was also applied to the GOODS-N Paschen-$\alpha$ sample). However, this is just one source with quite high attenuation ($A_V^{\rm neb}(\rm Pa\alpha H\alpha)=3.1\pm 0.1$ and 
$A_V^{\rm neb}(\rm BAGPIES)=3.8\pm 0.7$) so it is difficult to derive substantive conclusions.}. Note, if we assume $A_V^{\rm neb}=A_V^{\rm cont}/0.44$, then the mean correction factor becomes 1.07, but the final radio luminosity/SFR correlation in \S\ref{sec:result} is negligibly affected so the relationship between the stellar and gas attenuation does not have an effect on these results. The uncertainties on the  Paschen-$\alpha$ luminosities include both the measured flux uncertainty and that estimated for the $A_V$. 

We derive SFRs from the dust corrected Paschen-$\alpha$ luminosities. These SFRs are an instantaneous measurement as they arise from the nebulae directly illuminated by the most recently formed stars. We use the same conversion factor as \cite{Neufeld:24}: 
\begin{equation}
    \frac{\rm SFR_{\rm {Pa\alpha}}}
    {{\rm M}_\odot\,{\rm yr}^{-1}}\,=4.6\times 10^{-41}\times \frac{L_{\rm Pa\alpha}}{{\rm erg\,s}^{-1}}
    \label{eq:lpaa}
\end{equation}
which assumes case B recombination \citep[][]{Osterbrock:89}, an electron temperature of $10^4\,$K, a Chabrier initial mass function \citep[IMF,][]{Chabrier:03} and an intrinsic H$\alpha$ to Pa$\alpha$ ratio of 8.575  (this value can vary by $\sim\pm 10\%$ depending on electron temperature). This calibration ultimately comes from \cite{murphy:11} which is an updated version of the \cite{Kennicutt:98} nebula line calibration. Those papers assume a continuos (i.e. constant) SFR and solar metallicity. While the SFRs are averaged over 100\,Myr the nebular emission is dominated by light from young massive stars, for continuous star formation, so effectively it only traces the most recent 10\,Myr. Using a \cite{Kroupa:01} IMF instead makes negligible difference to the calibration \citep{murphy:11}. These calibrations ultimately rely on the Starburst 99 stellar models \citep{Leitherer:99}.

As a consistency check, we compare the {\tt BAGPIPES} SFR (specifically the instantaneous SFR value which is effectively close to a 10\,Myr averaged value) and Paschen-$\alpha$ SFR. The SFR values have a one-to-one relationship, but the {\tt BAGPIPES} values are on average $\sim70\%$ higher which we put down to the differing stellar population models and different metallicities. We discuss the effect of this difference on our results in \S\ref{sec:res:evolution}.


\subsection{MIGHTEE \texorpdfstring{$1.23\,$}GHz Continuum Data}
\label{sec:data:radio}

The MeerKAT International GHz Tiered Extragalactic Exploration survey \citep[MIGHTEE,][]{jarvis:16}\footnote{\url{https://www.mighteesurvey.org/}} is a $20\,\deg^2$ survey of well-studied fields to $\mu$Jy sensitivity conducted by the MeerKAT radio telescope. These observations in GOODS-South were performed in L-band covering 856-1711\,MHz and are the continuum data from the LADUMA {HI survey} project \citep{blyth:16}\footnote{\url{https://science.uct.ac.za/laduma}}.  The processing was performed in a similar manner to \cite{Heywood:22} and is described in \cite{Hale:25}. 

We use the deep image of the Extended Chandra Deep Field South (E-CDFS) from the MIGHTEE Data Release 1 \cite[DR1,][]{Hale:25}. This image covers $1.5\,\deg^2$  encompassing the GOODS-South region. It was imaged at two different resolutions with restoring beams of 5.5 and 7.3\,arcsec full width at half maximum (FWHM). These images have median measured root-mean-square (RMS) sensitivities at the centre of the each field of 1.2 and 1.3\,$\mu$Jy/beam respectively. The different resolutions come from using different values of the \cite{briggs:95} robust parameter: 0.0 and -1.2 for the low- and high-resolution images, respectively. Also provided is an effective frequency image which is necessary due to the size of the MeerKAT primary beam varying across the wide frequency range ($\Delta\nu/\nu\sim 1$). The mean effective frequency in the sub-region of E-CDFS covered by FRESCO is $\sim 1.23\,$GHz.

A source catalog is produced by running {\tt Python Blob Detection and Source Finder} \citep[{\tt PyBDSF},][]{mohan:15}\footnote{\url{https://pybdsf.readthedocs.io/}} code on each image. A $3\sigma$ threshold is used to detect islands of emission to ensure a high completeness of extended and compact emission, but only $5\sigma$ sources were retained in the final catalog. {\tt PyBDSF} also produces RMS maps using a sliding box although the RMS may be elevated in certain regions due to natural source confusion due to the sensitivity of the image. See \cite{Hale:25} for more details. 

We use the 5.5\,arcsec image and catalog due to its superior resolution. This catalog has additionally been cross-matched to host galaxies from the IR using a combined process of both statistical association and visual identification (further details in Hale et al. in prep). First, each \texttt{PyBDSF} source was matched to a $K_S$-band selected catalog from the VISTA Deep Extragalactic Observations \citep[VIDEO;][]{Jarvis2013} survey. To prepare the $K_S$ band selected catalog, a 5$\sigma$ limiting magnitude of 23.7 was applied and a star/galaxy separation was applied to remove potential stars, using $g-i$ (from the Hyper Suprime Camera Subaru) and $J - K_S$ (from VIDEO) colour cuts. A likelihood ratio (LR) technique \citep[see e.g.][]{Sutherland1992, McAlpine:13,weston:18, Williams2019} was then used to match the radio sources to a host galaxy. A decision tree was then adopted to determine whether the LR was sufficient, or whether visual inspection was required. Reasons for needing visual inspection include, but are not limited to a source being large such that it might be a lobed AGN and the LR not passing a sufficient threshold (which balanced completeness and reliability). Those sources which needed visual inspection were passed to an internal MIGHTEE Zooniverse \citep{Zooniverse}\footnote{\url{https://www.zooniverse.org}} repository which provided radio contours and the $K_S$ band host galaxies overlaid on $i$ (HSC-SSP) and 3.6 $\mu$m \citep[from the Spitzer Extragalactic Representative Volume Survey, SERVS,][]{Mauduit2012}\footnote{\url{https://doi.org/10.26131/IRSA407}} images. Five people were required to visually classify a source and a host galaxy was adopted where 60\% of these classifications agreed. 

\subsection{X-ray data}

We use the deep 7\,Ms {\it Chandra} X-ray catalog \citep{Luo:17} which covers 484.2\,arcmin$^{-2}$ of the E-CDFS including the FRESCO region. This catalog includes 1008 sources that are detected in up to three X-ray bands: $0.5-7.0$\,keV, $0.5-2.0\,$keV, and $2-7$\,keV. In the center of the image sensitivities of $1.9\times 10^{-17}$, $6.4\times 10^{-18}$, and $2.7\times 10^{-17}\,$ erg\,cm$^{-2}$\,s$^{-1}$ are reached in the three bands, respectively. 

This 7\,Ms catalog is cross-matched with the Paschen-$\alpha$ catalog using a one arcsec search radius (as that catalog has positional uncertainties up to $1\,$arcsec). At the sensitivity of this catalog it is possible to probe X-ray luminosities an order of magnitude below the rest-frame X-ray hard-band luminosity threshold for AGN, $L_{2-10keV}>3\times 10^{42}\,$erg\,s$^{-1}$, for the redshift range of this sample ($1.0\lesssim z \lesssim 1.8$).

\subsection{{\it JWST}/MIRI data}
Approximately half of the GOODS-S FRESCO survey area is covered by Systematic Mid-infrared Instrument Legacy Extragalactic Survey \citep[SMILES,][]{rieke:17}. SMILES covers $\sim 34\,$arcmin$^2$ ($3\times 5$ pointings) with eight bands across $\lambda\sim 5-27\,\mu$m. The $5\sigma$ point source sensitivities range from $0.21-15\,\mu$Jy from the shortest to longest wavelength, significantly deeper than the previous mid-IR data from {\it Spitzer} \citep{Dickinson:03a, Mauduit2012}. \cite{Lyu:24} use these data in conjunction with the deep NIRCam imaging from the {\it JWST} Advanced Deep Extragalactic Survey \citep[JADES,][]{Eisenstein:23} to model the SEDs of 3,273 MIR detected sources. This SED fitting identified 217 AGN candidates from their warm-dust components. Note our cross-identifications (see \S\ref{sec:agn:fresco}) include both confirmed AGN in their `massive galaxy' sample (stellar mass, M$_*\gtrsim 10^{9.5}\,$M$)\odot$) as well as candidate AGN in dwarf galaxies. In the latter case, the differing SEDs of dwarf galaxies combined with their lower luminosities makes robust AGN-detection via warm dust difficult. In this work, as we want to remove AGN, we take the conservative approach of considering the full sample of candidates as AGN. 

\subsection{Local Paschen-\texorpdfstring{$\alpha$} Sample}
\label{sec:data:local}
While ground based observations of local galaxies are complicated by atmospheric absorption, a sample of local ultra-luminous or luminous infrared galaxies (U/LIRGs) has been observed using a narrow band (NB) filter at $1.91\,\mu$m which coincides with an atmospheric window \citep[][]{tateuchi:15}. The observations were taken with the Atacama Near Infrared Camera \citep[ANIR,][]{Motohara:08} on the University of Tokyo Atacama Observatory \citep[TAO,][]{Yoshi:10} 1.0 m telescope \citep[miniTAO,][]{minezaki:10}. The sample of U/LIRGs was selected to be at Dec $<-30\,$deg and to have a small recessional velocity which would put the Paschen-$\alpha$ line emission in the $1.91\,\mu$m NB filter: $2800–8100\,$km\,s$^{-1}$ (corresponding to distances of $46.6–109.6\,$Mpc). 

Targets were selected from the {\it Infrared Astronomical Satellite} ({\it IRAS}) Revised Bright Galaxy Sample \citep[RBGS,][]{Sanders:03} and of the 151 RBGS targets matching the declination and velocity criteria above, 38 RBGS fields were randomly selected for observation. These fields included 44 galaxies in total since 12 galaxies formed six close pairs in the RBGS catalog. The galaxies in the sample were classified by their optical spectra as either HII (starforming), Seyfert 1/2 (signs of AGN) or LINER (high excitation state potentially associated with AGN activity). We cross-matched this sample with the NRAO VLA Sky Survey \citep[NVSS,][]{Condon:98} 1.4\,GHz radio catalog using a search radius of 45\,arcseconds (comparable to the NVSS resolution) finding 32 matches to the 38 fields. The six targets missed are due to the NVSS lower declination limit of $-40\,$deg. 

\begin{figure*}[ht!]
\includegraphics[trim={0.3cm 0.3cm 0.3cm 0.3cm},clip,width=0.99\linewidth]{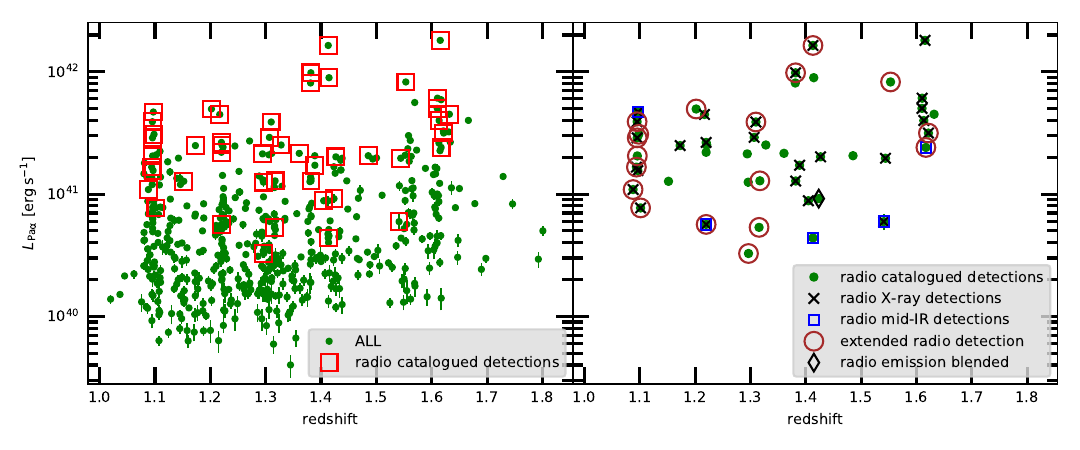}
\vskip -0.cm
\caption{Attenuation corrected line luminosities with uncertainties plotted against redshift. {\it (left)} The 506 Paschen-$\alpha$ sources in the GOODS-South sample with red squares indicating the 47 sources which match the host galaxies of the cataloged radio sources from the E-CDFS MIGHTEE image at $1.23\,$GHz. {\it (right)} Radio cataloged detections of Paschen-$\alpha$ sources with AGN identified (as described in \S\ref{sec:agn}). AGN are selected via X-ray detections from a deep {\it Chandra} image \citep{Luo:17}, mid-IR warm dust from SED fitting using NIRCam/MIRI data \citep{Lyu:24} or from extended radio emission from MIGHTEE (\S\ref{sec:data:radio}). One source is removed due to blended radio emission. Of 47 radio sources only 11 are SFGs free from AGN activity. 
\label{fig:zlpa}}
\end{figure*}

We use the Paschen-$\alpha$ luminosities reported by \cite{tateuchi:15} as they use the same cosmology. For the six close pairs we sum the luminosities as the relatively low resolution of NVSS can only provide flux densities for the total systems. The total extinction in these systems is determined from the Balmer decrement. \cite{tateuchi:15} then use the \cite{calzetti:00} extinction curve with $R_V=4.05$ to correct their Paschen-$\alpha$ SFRs for this extinction. We therefore correct their observed  Paschen-$\alpha$ luminosities by the ratio of the corrected SFR to the uncorrected SFR. Uncertainties on the final luminosities used here are a combination of the flux uncertainty as well as the uncertainty on the extinction correction.  

\section{Analysis} \label{sec:analysis}

\subsection{Cross-matching}

We cross-match the Paschen-$\alpha$ catalog with the MIGHTEE radio catalog using the positions of the identified host galaxies of the MIGHTEE radio sources rather than the radio positions. Using a search radius of 0.5\,arcsec we find 47 matches. We justify this radius due to the FRESCO and VIDEO\footnote{which define the host galaxy positions of the MIGHTEE radio sources.} data having positional accuracies of $<0.1\,$arcsec and the galaxies are all $<1\,$arcsec at this redshift. We also cross-match the Paschen-$\alpha$ catalog with the mid-IR SMILE catalog using the same 0.5\,arcsec radius. In the left panel of Fig.~\ref{fig:zlpa} we plot the Paschen-$\alpha$ line luminosities of the full sample as a function of redshift and indicate those sources with MIGHTEE 1.23\,GHz detections. 

\subsection{Identifying AGN}
\label{sec:agn}
\subsubsection{FRESCO \texorpdfstring{$z\sim 1.3$} Sample}
\label{sec:agn:fresco}
As this analysis is focused on the radio properties of Paschen-$\alpha$ selected SFGs we wish to filter out AGN. Of the 506 FRESCO sources from \S\ref{sec:data:fresco}, 50 are found to have an X-ray counterpart including 26 of the 47 radio detections. While the X-ray luminosities of these sources are below the canonical maximum value used to securely identify AGN \citep[$L_X\ge 3\times 10^{42}\,$erg\,s$^{-1}$,][]{Luo:17}, they are above the luminosity expected for their SFR \cite[e.g.][]{Symeonidis:11} so we conservatively flag them as AGN (although some could be extremely dusty SFGs).

We perform similar removal of AGN using the SMILE catalog of mid-IR selected AGN. Of the 506 FRESCO sources 17 are identified as mid-IR AGN including five sources with radio-detections. The SMILE survey only covers around half of the FRESCO survey so this mid-IR AGN selection is incomplete. 

As any SFGs in the final Paschen-$\alpha$ sample are expected to be less than one arcsec in size, we can remove extended radio-loud AGN by flagging all radio sources with a non-zero deconvolved size in the image plane. This approach potentially removes some SFGs at low signal-to-noise ratio (SNR), but we would rather lose a few SFGs and minimize the AGN contamination. As the radio resolution of 5.5\,arsec is roughly an order of magnitude larger than the typical sizes of SFGs at this redshift \citep[e.g.][]{Vanderwel:14} this selection is a good way to remove radio sources powered by jets and lobes. This selection flags 18 sources. The two Paschen-$\alpha$ sources with broad lines, mentioned in \S\ref{sec:data:fresco}, are both detected in the X-ray so are excluded by the first step above (one is also a radio detection). We also perform visual inspection to ensure robust cross-matching and remove a further source due to having another galaxy contribute to the measured radio flux density. 

The combined visual/X-ray/mid-IR/radio selections above leave just 11 non-AGN radio detections in the final Paschen-$\alpha$ sample. We indicate the sources identified as AGN in the right panel of Fig.~\ref{fig:zlpa}. Table~\ref{tab:numbers} provides a running count of the Paschen-$\alpha$ sample following these and subsequent selection criteria. 

To determine the radio luminosities accurately we need the spectral indices. As these are unknown we assume a spectral index (evaluated above 1\,GHz) of $\alpha=-0.78\pm 0.15$ which is the mean and standard deviation of the spectral indices from the local sample of SFGs studied by \cite{grundy:25}. Using a lower spectral index of $\alpha=-0.5$, i.e. with more free-free contribution, only decreases the luminosities by $\sim 20\%$. To obtain the 1.4\,GHz luminosities we first convert the 1.23\,GHz flux densities to a 1.4\,GHz flux densities, $S_{\rm 1.4GHz}$, using this spectral index. The 1.4\,GHz radio luminosity, $L_{\rm 1.4GHz}$, is therefore: 

\begin{equation}
    \frac{L_{\rm 1.4GHz}}{{\rm W\,Hz}^{-1}}=4\pi\,\bigg(\frac{d_{\rm L}}{\rm m}\bigg)^2\,\frac{S_{\rm 1.4GHz}}{\rm 10^{-26}\,Jy}\,(1+z)^{-1-\alpha},
    \label{eq:lrad}
\end{equation}

\noindent
where $d_{\rm L}$ is the luminosity distance at a redshift, $z$ and $S_{\rm 1.4GHz}$ is the 1.4\,GHz observed-frame flux density. The uncertainties on the radio luminosities are propagated from the flux and spectral index uncertainties as we assume all redshifts have negligible uncertainties.

\begin{deluxetable}{lcc}[t!]
\tablecaption{Breakdown of the 506 Paschen-$\alpha$ sample following cross-matching with the radio catalog and cuts described in \S\ref{sec:agn:fresco}, \S\ref{sec:radio:blended} and \S\ref{sec:stacking}. Numbers in brackets represent subsamples to be subtracted from the number above. The final breakdown of non-AGN sources (in bold below) is: 11 cataloged radio detections, 18 non-cataloged radio detections and 298 non-detections to be stacked.}
\label{tab:numbers}
\tablehead{
\colhead{Description} & \colhead{Non-detections} & \colhead{Detections} 
}
\startdata
full sample & 459 & 47 \\ \hline 
X-ray AGN & (24) & (26) \\ \hline
radio AGN & & (18) \\ \hline
mid-IR AGN & & (5) \\ \hline
visual inspection & & (1) \\ \hline
cataloged radio detections & & 11 $^a$\\ \hline 
blended with radio & (118) & \\ \hline
remaining X-ray AGN$^b$ & (17) & \\ \hline
remaining mid-IR AGN$^b$ & (10) & \\ \hline
non-cataloged detections & (18) & \\ \hline 
non-detections & 298$^c$ & \\ \hline 
\enddata
\tablecomments{$^a$the X-ray/radio/mid-IR AGN sub-samples overlap making a total of 35 AGN, $^b$after removing blended sources, $^c$the X-ray and mid-IR AGN sub-samples overlap making a total of 25.}
\end{deluxetable}

\subsubsection{Local \texorpdfstring{$z<0.1$} Sample}
\label{analysis:local}

\begin{figure}[t!]
\includegraphics[trim={0.3cm 0.3cm 0.3cm 0.3cm},clip,width=0.99\linewidth]{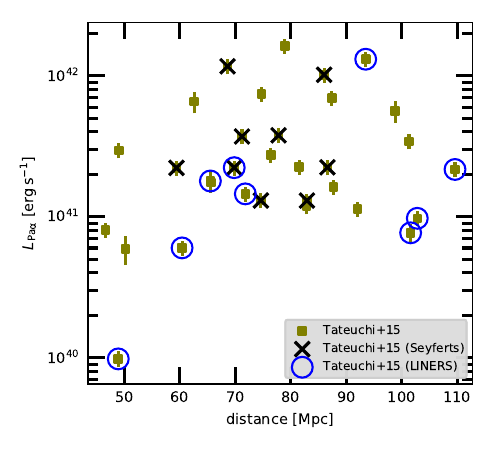}
\caption{Distribution in Paschen-$\alpha$ luminosity/distance parameter space of the local sample from \cite{tateuchi:15} as presented in \S\ref{sec:data:local}. Sources with Seyfert or LINER indicators in their optical spectrum are marked. 
These are removed from further analysis to avoid contamination by AGN as discussed in \S\ref{analysis:local}. 
\label{fig:local}}
\end{figure}

From visual inspection, the radio emission of all the local sources appears to be related to star formation, i.e. no signs of jets or lobes are apparent, with the emission broadly tracing the underlying optical images. 
The spectral classification provided by \cite{tateuchi:15} allows us to identify potential AGN in the local sample. We remove sources clearly identified as AGN from a Seyfert classification (nine sources) and, out of an abundance of caution, also remove those with a LINER classification (nine sources) as the high excitation state of LINERs could be due to an AGN. Hence, in Fig.~\ref{fig:local} we plot the Paschen-$\alpha$ luminosities as function of distance of the local sample indicating which sources are classified as Seyferts or LINERS. In subsequent analysis we remove these two AGN subsets leaving just 14/32 local sources. Radio luminosities are determined via Equation~\ref{eq:lrad}, but note that the $k$-correction is negligible at very low redshift.

\subsection{FRESCO Paschen-\texorpdfstring{$\alpha$} Sources Blended with Radio}
\label{sec:radio:blended}
The lower resolution of the radio image, by about an order of magnitude, compared to the typical size of SFGs found by the Paschen-$\alpha$ detections at this redshift presents a challenge for the stacking. We do not want to stack Paschen-$\alpha$ sources which have nearby, but unrelated radio emission, hence
we remove all non-radio detected Paschen-$\alpha$ sources which lie within $7\,$arcsec of a cataloged radio source. We find we have to use a $7\,$arcsec radius, larger than the FWHM of the radio image ($5.5\,$arcsec), in order to be sure that none of the remaining non-detected Paschen-$\alpha$ sources have radio flux densities contaminated by nearby sources. This cut finds a total of 118 potentially blended sources (including three from visual inspection). 
The blended Paschen-$\alpha$ sources include seven with X-ray detections so the number of Paschen-$\alpha$ sources without radio detections, but with X-ray detections, falls to 17. 

\subsection{Stacking of FRESCO Paschen-\texorpdfstring{$\alpha$} Sources}
\label{sec:stacking}
To obtain the mean radio flux density of the Paschen-$\alpha$ sources without radio detections we can stack in the radio image. Given the 5.5\,arcsec resolution of MIGHTEE, all the FRESCO Paschen-$\alpha$ SFGs will be unresolved so the flux from a single pixel provides a good measure of the total flux. 

\subsubsection{Background Subtraction and Confusion Noise}

When stacking at the positions of the Paschen-$\alpha$ sources we have to subtract the background in the radio image. We take all the pixels in the FRESCO sub-region of the ECDFS MIGHTEE image and fit a Gaussian to just the negative side of a histogram of their values. We then equate the background to being the mean of this Gaussian. From this method we obtain a local background of $-0.97\,\mu$Jy which we subtract from the flux density at each pixel. 

We also have to account for confusion noise since the radio images are approaching the confusion limit. To determine the amount of flux contributing to each pixel from nearby non-associated radio sources, we take 10,000 realisations of 298 random positions (i.e the same number we stack later) and measure the median flux each time. We then take the mean of these 10,000 medians to estimate the confusion noise to be subtracted. From this method we obtain a confusion noise of $0.46\pm 0.17\,\mu$Jy/beam with the uncertainty being the standard deviation of the 10,000 realisations. 

So for the corrected single pixel flux measurements we subtract our estimates of both the background in the FRESCO region and the confusion noise. We note that this method accounts for the confusion noise statistically rather than on an individual basis, but is appropriate given the sample size and depth.

\subsubsection{Non-cataloged detections}

\begin{figure}[t!]
\includegraphics[trim={0.3cm 0.3cm 0.3cm 0.3cm},clip,width=0.99\linewidth]{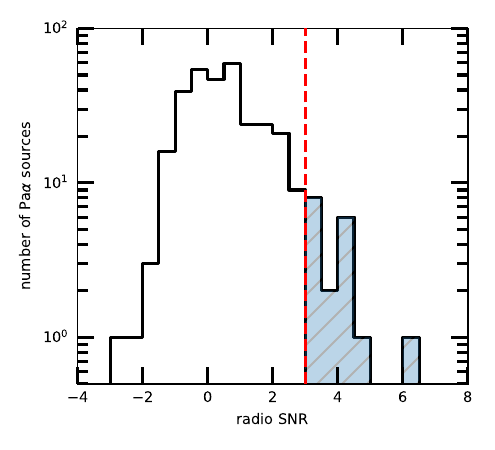}
\caption{Histogram of the SNR of MIGHTEE flux density pixels values at the positions of Paschen-$\alpha$ sources without counterparts in the radio catalog. The 18 sources with values above $3\sigma$ (the vertical dashed line) are considered non-cataloged detections. 
\label{fig:hist}}
\end{figure}

Using the local RMS image we can determine the signal-to-noise ratio (SNR) of these sources without counterparts in the radio catalog using the corrected flux measurement described above. We plot the distribution of the SNR in Fig.~\ref{fig:hist} and find a tail of sources with positive SNRs which extends to $5\sigma$, but with one source at $6.0\sigma$. Above $3\sigma$ we find 18 sources which we do not stack, but treat as `non-cataloged radio detections'. We determined their radio luminosities and uncertainties in the same manner as the cataloged radio detections. 

The sole $>5\sigma$ detection is likely due to the complication of using a source finder on a crowded, confusion limited survey where estimating the local RMS is complex. Visual inspection suggests this is real radio flux likely associated with the Paschen-$\alpha$ emitter missed by {\tt PyBDSF}. Hence, we include the $6.0\sigma$ non-cataloged radio detection going forward, but its inclusion does not materially affect the results. 

\subsubsection{Paschen-\texorpdfstring{$\alpha$}~~Sources Blended with  each other}
The low resolution of the MIGHTEE radio data also means that some of these non-detections and non-cataloged detections will be close enough together to be within the same resolution element of the radio image. We check for close pairs using a 5.5\,arcsecond radius and find three and 64 close pairs from the 18 non-cataloged detections and the 298 remaining non-detections samples, respectively. In all cases, these pairs are far enough apart that they have different radio flux densities from separate, albeit very close, pixels. However these nearby pixel-based radio fluxes are not independent measurements when there are two Paschen-$\alpha$ sources  within one radio resolution element. 

We make the assumption that the two Paschen-$\alpha$ sources both contribute to the radio flux and that the total radio flux is the higher of the two pixel values. Hence, we take the higher flux density value and assign a portion of the radio emission to each of the pair based on their relative Paschen-$\alpha$ flux (from the assumption that the two values are indeed correlated)\footnote{Note, assigning this radio flux equally to the two sources in a pair does not substantially affect the results.}. For the stacked sample this is equivalent to assigning the higher flux density to just one source. Sharing the flux by the Paschen-$\alpha$ flux ratio affects six of the 18 non-cataloged radio detections by decreasing their radio luminosity. 

\subsubsection{Stacking Method}

The final stacked flux density is simply the median of the MIGHTEE pixel values, corrected for background and confusion noise, at the positions of each of the 298 Paschen-$\alpha$ sources (including the cases where we re-evaluated the flux densities of close pairs of sources). We find a median value of $0.56\,\mu$Jy which is a little lower than the mean, $0.59\,\mu$Jy. The mean being close to the median suggests that there is not a significant bright tail of sources (potentially radio-excess AGN) left in the stack.

To estimate the uncertainty on this stacked flux we repeat the stacking process at 298 random positions 10,000 times. We then take the standard deviation of these 10,000 trials as the uncertainty on the stacked flux density: $0.18\,\mu$Jy which provides a $3.3\sigma$ stacked flux density value. 

The radio luminosity of the stack is determined from the stacked flux using Equation~\ref{eq:lrad} using the mean redshift of the 298 sources ($\langle z\rangle = 1.31$). We obtain a radio luminosity of $L_{\rm 1.4GHz}=5.5\pm1.9 \times 10^{21}\,$W\,Hz{\bm{$^{-1}$}} where the uncertainty is the propagated combination of the flux uncertainty and that of the spectral index. We note that we could have determined the radio luminosity of each of the 298 non-detections separately (using the individual pixel flux densities) and then taken the mean. Doing so, in fact, gives a very similar value: $L_{\rm 1.4GHz}=5.6 \times 10^{21}\,$W\,Hz$^{-1}$.

\subsection{Modelling}

\subsubsection{Calibrating the Radio/SFR Conversion at \texorpdfstring{$z\sim 1.3$}~}

\label{sec:mod:conv}
From our final FRESCO sample of AGN-free SFGs (11 cataloged radio detections, 18 non-cataloged radio detections and the 298 non-detections) we have an opportunity to investigate the radio luminosity/SFR correlation for a well-defined sample with well-determined SFRs at $z\sim 1.3$. We use the python routine {\tt linmix} \citep{kelly:07} which uses a hierarchical Bayesian approach to linear regression accounting for uncertainties on the both axes as well as non-detections (as we have here). {\tt linmix} runs a Markov Chain Monte Carlo, randomly sampling from the posterior, to produce samples from the posterior distribution of the model parameters, given the data. The {\tt linmix} algorithm is based on a model which derives a likelihood function for the measured data and assumes that the distribution of two variables can be approximated using a mixture of Gaussians. In the case of non-detections or limits, the likelihood function is modified to include this selection effect. We fit the data in log-space so that the linear relation provides a power law. For the undetected sources we use their $3\sigma$ flux densities as upper limits. 

\subsubsection{Modelling High (and Low) Radio Luminosities}
\label{sec:mod:evolution}
\begin{figure}[t!]
\includegraphics[trim={0.3cm 0.3cm 0.3cm 0.3cm},clip,width=0.99\linewidth]{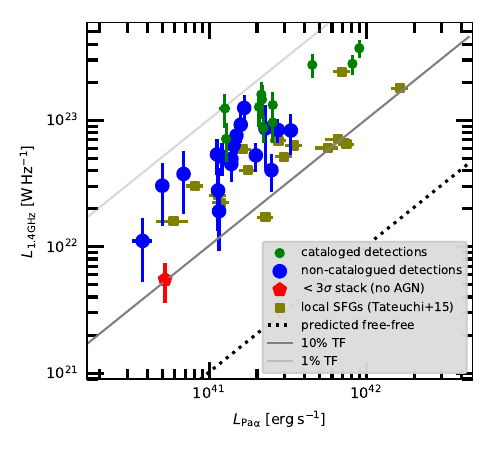}
\caption{Radio luminosity plotted as function of Paschen-$\alpha$ luminosity for our various AGN removed sub-samples: the 11 cataloged radio-detections, the 18 non-cataloged detections and the stack of the 298 remaining non-detections after removing blended sources. The local sample of 14 Paschen-$\alpha$ emitters from \cite{tateuchi:15}, minus AGN, is also plotted. Both axes have error bars plotted which are sometimes smaller than the symbol. Overlaid as lines are the expected 1.4\,GHz free-free luminosity from a Paschen-$\alpha$ source as well as $\times 10$ and $\times 100$ this value which corresponds to 10\% and 1\% 1.4\,GHz thermal fractions (TFs), respectively. Nearly all galaxies have TF  in the range $1-10\%$ with the local sample having slightly higher TF.
\label{fig:lumlum}}
\end{figure}

Fig.~\ref{fig:lumlum} shows a number of SFGs with relatively high radio luminosities for their SFR. i.e. very low thermal fractions $\le 5\%$. We investigate whether at least some of the scatter of sources around these correlations could be due to the radio luminosity being a delayed probe of star formation. Recent work \citep[][]{cook:24} suggest that radio emission in SFGs correlates better with the SFR a few hundred Myr in the past (as inferred from star-formation histories derived from SED modelling) than with tracers of the current SFR. 

The relativistic electrons, cosmic rays (CR), responsible for the synchrotron emission are produced at the end of the lives of the most massive stars which can live for $10-30\,$Myr for  type II supernova progenitors. The timescale of the synchrotron emission depends on the lifetime of the CRs and there is growing evidence that the radio emission is a delayed tracer of SFR by potentially up to a few hundred Myr as suggested by \cite{cook:24}. Estimates of the total lifetime of these CR vary as there are two contributions: the synchrotron cooling time and the escape time from the galaxy. The former has estimates of $10-100\,$Myr \citep{heesen:23} and the latter can be $20-300\,$Myr \citep{dorner:23}. Combining these results gives total lifetimes ranging across $7-75\,$Myr.

Tracers of the SFR which come directly from the interstellar medium in star forming regions include the free-free radio emission and the hydrogen recombination lines such as Paschen-$\alpha$. These are expected to be near instantaneous (delayed by just 10\,Myr and averaged over the last 10\,Myr). 

Here we create a toy model where we follow the recent SFR (as traced by Paschen-$\alpha$) and the total radio emission (free-free plus synchrotron emission) through a starburst phase, but with the synchrotron emission delayed by $20\,$Myrs (the mean time for the massive stars to go supernova) and then averaged over the previous 75\,Myr, the upper limit of the potential range. The delay and average timescales for all components are summarised in Table~\ref{tab:timescales}.

In order to follow our toy model during an increasing and decreasing starburst phase we use a Cauchy-Lorentz function for the star-formation history as it has an approximate exponential rise and fall timescale. This model has a base SFR of 3\,M$_\odot$\,yr$^{-1}$ and a peak SFR of 30\,M$_\odot$\,yr$^{-1}$ with a time scale ($\gamma$ in the Cauchy-Lorentz function) of 100\,Myr. The relative SFRs are chosen to represent the likely extremes of starbursting and quenching, but the absolute values are arbitrary and chosen to fit within the distribution of observed Paschen-$\alpha$ sources. While rapid changes are uncommon in the local Universe, they may not be at high redshift as seen in hydrodynamical simulations \citep[e.g.][]{muratov:15} although the timescale here, $100\,$Myr is quite rapid. We discuss how these choices affect the results in \S\ref{sec:res:evolution}. From this model we trace the instantaneous SFR as well as the delayed radio emission over time. 
\setlength{\tabcolsep}{20pt}
\begin{deluxetable}{lcc}[t!]
\tablecaption{Timescales for the delay (relative to the star formation) and for the average over the prior preriod for the toy starburst model.}
\label{tab:timescales}
\tablehead{
\colhead{Component} & \colhead{Timescale} & \colhead{Duration} 
}
\startdata
\multirow{2}{*}{Pa-$\alpha$} & delay & 10\,Myr\\
            & average & 10\,Myr\\
\hline
\multirow{2}{*}{free-free}   & delay   & 10\,Myr\\
            & average & 10\,Myr\\
\hline
\multirow{2}{*}{synchrotron} & delay   & 20\,Myr\\
            & average & 75\,Myr\\
\enddata
\end{deluxetable}

As the radio emission is a combination of synchrotron and free-free, we assume an initial thermal fraction ($TF$) of $5\%$ at 1.4\,GHz \citep[$TF_{\rm 1.4}$ as typically seen in local starbursts,][]{Galvin:18,grundy:25}. We can trace the evolution of the observed $TF_{\rm 1.4}$. We use the SFR to radio luminosity conversion factor determined in the previous section to convert the delayed SFR into a radio luminosity. However, our choice of conversion factor is not important as this only normalises the evolutionary tracks and we are interested in the deviation from the mean value. 

\begin{figure*}[t!]
\centering
\includegraphics[trim={0.3cm 0.3cm 0.3cm 0.3cm},clip]{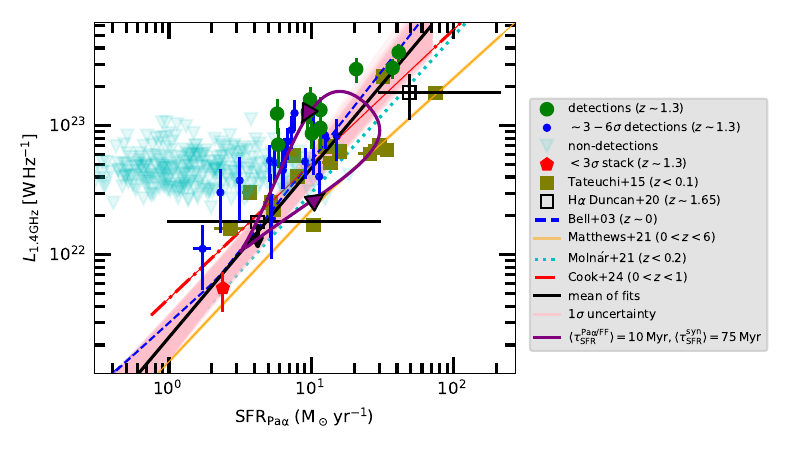}
\caption{Radio luminosity plotted as function of Paschen-$\alpha$ SFR for our various sub-samples after the removal of AGN: the 11 cataloged radio detections, the 18 non-cataloged radio detections, the 14 local sources and the 298 remaining Paschen-$\alpha$ non-detections. We also present the stacked value of these non-detections although it is not used in the fitting. For comparison we present the H$\alpha$/radio luminosity results from \cite{duncan:20} at $z\sim 1.65$. 
We overlay various conversion factors from the literature including \cite{Bell:03,molnar:21,cook:24} and \citet{matthews:21} as well as indicating the redshift range they were evaluated over (the length of the line indicates the SFR range they were determined over). The best fit radio luminosity/SFR correlation is presented (see \S\ref{sec:mod:conv}) with the pink tracks representing solutions within  $1\sigma$ of the median fit of both parameters. We also overlay a starburst evolutionary track as described in \S\ref{sec:mod:evolution} with the arrows indicating the direction of time. The starburst has a Cauchy-Lorentz (pseudo-exponential) rise and fall temporal form with the synchrotron component of the radio luminosity delayed by 20\,Myr (compared to 10\,Myr for the Paschen-$\alpha$ and free-fee components). The SFRs are averaged over previous 10 and 75\,Myr for the Paschen-$\alpha$ SFR/free-free luminosity and synchrotron luminosity, respectively. This model can partly explain the scatter around the best fit line in this and other work. 
\label{fig:lumsfr}}
\end{figure*}

\section{Results and Discussion} \label{sec:result}

\subsection{Radio v Paschen-\texorpdfstring{$\alpha$} Luminosities}

In Fig.~\ref{fig:lumlum} we plot the radio luminosity against the Paschen-$\alpha$ luminosity of the detected radio sources, the non-cataloged radio detections, the local sample as well as the stacked value. The cataloged sources are naturally above the non-cataloged radio detections which in turn are above the stacked limit. This distribution is clearly due to the sensitivity of the radio survey which applies a roughly constant luminosity cut for this redshift range. The stacked value clearly accounts for the radio luminosities of the non-detected radio sources which undoubtedly extend down to $10^{21}\,$W\,Hz$^{-1}$ and below. 

In Fig.~\ref{fig:lumlum} we also present the predicted free-free 1.4\,GHz luminosity for a given Paschen-$\alpha$ luminosity using: 

\begin{equation}
\frac{L_{\rm ff}(\nu)}{\rm W\,Hz^{-1}}=1.062\times 10^{-20}\,\bigg(\frac{T}{10^4\,{\rm K}}\bigg)^{0.45}\,\bigg(\frac{\nu}{\rm GHz}\bigg)^{-0.1}\,\frac{L_{\rm Pa\alpha}}{\rm erg\,s^{-1}},
\end{equation}

\noindent
where $L_{\rm ff}(\nu)$ is the free-free luminosity as a function of frequency, $T$ is the temperature of the ISM (which we take to be $10^4\,$K as before). This equation comes from combining the relationship of the photo-ionization rate and thermal radio luminosity \citep[][]{rubin:68} and that between the photo-ionization rate and Paschen-$\alpha$ luminosity \citep[][]{Kennicutt:98}. Combining this equation with Equation~\ref{eq:lpaa} gives a result within $10\%$ of other free-free/SFR relations used in the literature \citep[e.g.][who used a different IMF]{murphy:11}.

It is interesting to note the range of thermal fractions of this sample at 1.4\,GHz, $TF_{\rm 1.4GHz}$. Generally for low-z SFGs the $TF_{\rm 1.4GHz}$ varies from around $1\%$ to $10-30\%$ in some extreme cases \citep[][]{Galvin:18,grundy:25}. The local sample have $TF_{\rm 1.4GHz}\approx 2-20\%$ whereas the $z\sim 1.3$ detected Paschen-$\alpha$ sample (cataloged and non-cataloged) range from $\approx 1-10\%$. We note that the lack of high $TF_{\rm 1.4GHz}$ in the detected sample is due to selection effects, i.e. we are biased against directly detecting sources with low radio luminosities for their Paschen-$\alpha$ luminosity. We investigate if the low $TF_{\rm 1.4GHz}$ sources are fading starbursts where the radio emission is still high due to the delayed timescales of synchrotron emission using the model described in \S\ref{sec:mod:evolution}. We show these models in Fig.~\ref{fig:lumsfr} and discuss them in \S\ref{sec:res:evolution}.

\subsection{Radio Luminosity vs SFR}

In Fig.~\ref{fig:lumsfr} we plot the radio luminosity against the Paschen-$\alpha$ derived SFR overlaid with four different 1.4\,GHz luminosity to SFR conversion factors \citep[][all with a Chabrier IMF]{Bell:03,molnar:21,cook:24,matthews:21}. While there are numerous more relations in the literature these four are chosen to demonstrate the breadth of relations in use. Each is determined over a different redshift range as indicated in the legend. If one includes the stacked data point representing 298 SFGs we can see that the data here are consistent with these different relations albeit with a scatter at high radio luminosities which we discuss in \S\ref{sec:res:evolution}. The local sample also agrees well with these previous relations. 

\subsubsection{Correlation and Fit}
\label{sec:res:corr}

The result of the power law fit (using {\tt linmix}) to the data at $z\sim 1.3$ is:
\begin{equation}
     L_{\rm 1.4GHz}[{\rm W\,Hz}^{-1}]\!\!=\!\!10^{21.36\pm 0.17}\times{\rm SFR_{Pa\alpha}}[{\rm M}_\odot\,{\rm yr}^{-1}]^{1.31\pm 0.17},
\end{equation}

\noindent
where the values are the medians and the uncertainties the 68.2 ($1\sigma$) percentile limits. We plot this fit in Fig.~\ref{fig:lumsfr}. For reference we overlay all chain models from the {\tt linmix} fitting that have parameter values within $1\sigma$ of the best fit values as indicated by the pink region. 

This fit is consistent with most of the previous relations plotted \citep[i.e.,][]{Bell:03,molnar:21,cook:24}, demonstrating that this Paschen-$\alpha$-selected sample follows other commonly derived 1.4\,GHz luminosity to SFR conversions. We note that our relation and others lie around $0.5\,$dex above the \citet[][hereafter M21]{matthews:21} relation at the high SFR end (i.e. $50-100\,$M$_\odot$\,yr$^{-1}$). The M21 result comes from a fit to the source counts and a P(D) analysis down to $\sim 0.2\,\mu$Jy so it is probing the brightest radio sources across all redshifts. Hence, part of the fit comes from the luminous SFG at $z>3$ where iC losses may become important \citep[e.g.,][]{whittam:24}. The radio luminosity/SFR conversion from M21 does not account a potential variation in redshift so we postulate that difference seen in Fig.~\ref{fig:lumsfr} could be due to unaccounted for iC losses. Direct calibration of the radio luminosity/SFR conversion at $z>3$ would confirm this or not.

Our result is also agrees well observational data at $z\sim 1.65$ from \cite{duncan:20} although their high SFR datum is marginally lower. The fit also broadly agrees with the local sample. In order to improve our fit we would need some combination of more direct detections, a larger sample, probing higher and lower SFRs. We note our result is not driven by systematics, for example changes to the dust attenuation or the radio k-corrections only shift luminosities by up to $0.1-0.2$ dex. Also, if we increase the confusion noise from $0.45\,\mu$Jy/beam to $0.8\,\mu$Jy/beam, a few sources switch from non-cataloged detections to non-detections, but the overall fit remains the same to one decimal place.

\subsubsection{High/low Radio Luminosities}
\label{sec:res:evolution}

As described in \S\ref{sec:mod:evolution} we have developed a toy model to explain the SFGs with relatively high or low radio luminosities for their SFRs/Paschen-$\alpha$ luminosities. This model assumes that the dominant synchrotron emission is delayed by $10\,$Myr relative to the Paschen-$\alpha$ and free-free emission, and that the Paschen-$\alpha$/free-free and synchrotron emission are averaged over 10 and $75\,$Myr respectively. We present this evolutionary track in Fig.~\ref{fig:lumsfr}. 

The track initially increases more rapidly in SFR$_{\rm Pa\alpha}$ than $L_{\rm 1.4GHz}$  with the SFR$_{\rm Pa\alpha}$ peaking and then starting to decline before the $L_{\rm 1.4GHz}$ peaks. The $L_{\rm 1.4GHz}$ then peaks before declining more rapidly than SFR$_{\rm Pa\alpha}$ for the track to end up at the start. The model shows that a delay in the synchrotron emission can cause a scatter of almost one dex, covering a large part of the range of sources lying on this correlation. Due to the radio detection limit we do not directly sample the full range of scatter below the fitted relation for the Paschen-$\alpha$ sample. Using a shorter or longer delay or time average would cause a narrower or wider scatter, respectively although we note that the width of the scatter is driven more by the time averaging unless the delay time becomes comparable to the average time. This toy delayed/averaged synchrotron model can potentially explain some of the scatter observed in the radio luminosity/Paschen-$\alpha$ SFR relation and indeed any relation between radio luminosity and a more instantaneous measure of SFR (e.g. the radio/far-infrared relation). 

We caution the reader that there may be other effects, on top of measurement error, which could contribute to the scatter. Uncertainty in the dust correction for the Paschen-$\alpha$ sources leads to shifts of $\sim 0.1\,$dex. The choice of the IMF has negligible effect on the Paschen-$\alpha$-derived SFR \citep{murphy:11} although the ionizing spectrum hardness may have a stronger effect. Other factors affect the SFR$_{\rm Pa\alpha}$ estimate including the underlying choice of stellar populations and metallicity. As an example the $\sim 70\%$ offset with the {\tt BAGPIPES} SFR shows that systematic shifts can occur. However, systematic shifts simply change the normalisation in Fig.~\ref{fig:lumsfr}. The radio emission also depends on the total magnetic field of the galaxy however, this is found to vary by only a factor of a few \citep{Tabatabaei:17, Tabatabaei:25}. It is hard to see how effects like varying geometries and uneven dust distribution can substantially add to the scatter. Potentially changes to the synchrotron spectrum from synchrotron or ionisation losses could contribute, but this is expected to only become dominant at very high redshifts \citep[$z>3$,][]{whittam:25}. 

While the underlying thermal fraction may also be a source of systematic uncertainty, this model can also explain the wide range of TF observed in local galaxies \citep[e.g.,][]{Galvin:18,Dey:22,Dey:24} since the free-free contribution rises and falls before the synchrotron component. The TF reaches a minimum of $\sim 1\%$ and maximum of $\sim 20\%$ at the furthest distances away from the best fit relation. Hence, the TF could act as tracer of the age of a starburst in extreme situations. e.g. the very high thermal fraction seen in Haro 11 \citep[][Grundy et al., in prep]{Komarova:24}. 

A prediction of this model is that young starbursts with rising SFRs would lie below the correlation, have higher thermal fractions and flatter spectral indices. Old starbursts with declining SFRs on the other hand would lie above the correlation with low TF and steeper spectra. Upcoming MIGHTEE S-band ($2-4\,$GHz) data may potentially be able to test the variation in spectral index above/below the radio/SFR. Furthermore, in such a scenario, outliers above/below the radio/SFR relation would feature distinct star formation change parameters \citep{wang:20}. These predictions will be investigated in future work. 

\section{Conclusions} \label{sec:conclusion}

We have investigated the radio properties of a sample of {\it JWST}/FRESCO Paschen-$\alpha$ selected sources across $1\lesssim z\lesssim 1.8$ in order to determine if the SFGs at this epoch follow previously derived radio luminosity-SFR conversions. Our main results are:

\begin{enumerate}
    \item Of 506 FRESCO Paschen-$\alpha$ sources in the GOODS-South, 47 match with the hosts of cataloged radio sources at $1.23\,$GHz from the MIGHTEE ECDF-S image. 
    \item Removing likely AGN and blended sources leaves a sample of SFGs comprised of: 11 cataloged radio detections, 18 non-cataloged radio detections and 298 undetected sources. 
    \item Stacking the undetected sources in flux space we obtain a $3.3\sigma$ detection. 
    \item The full FRESCO Paschen-$\alpha$ sample spans approximately two orders of magnitude in both Paschen-$\alpha$ and radio luminosity. Overlaying this sample, along with a local sample of Paschen-$\alpha$ SFGs, we find they follow existing correlations of radio luminosity and SFR across a range of redshifts from the literature. 
    \item Specifically, by fitting our $z\sim 1.3$ sources (cataloged detections, non-cataloged detections and non-detections) we find a relation of $\log(L_{\rm 1.4GHz})= (1.31\pm0.17)\times \log({\rm SFR}_{\rm Pa\alpha}) + (21.36\pm 0.17)$ which is consistent with the literature values.
    \item Some of the width of the scatter in the $L_{\rm 1.4GHz}$/SFR$_{\rm Pa\alpha}$ correlation can be explained by synchrotron emission being a delayed/averaged tracer of star formation by $10/75\,$Myr. 
\end{enumerate}

We have confirmed that this sample of SFGs does have radio luminosities consistent with previous calibrations of the $L_{\rm 1.4GHz}$/SFR relation. We can explain some of the scatter in this relation as well, but more sophisticated models with more realistic star-formation histories are needed to elaborate upon the toy model presented here as well as more direct detections of high-z SFGs in the radio. However, we have demonstrated that this concept is a feasible explanation for a good part of the scatter. Furthermore, this model provides predictions for the nature of the radio emission for rising and declining starbursts which can be tested in the future.

\begin{acknowledgements}
We acknowledge the Noongar people as the traditional owners and custodians of Whadjuk Boodjar, the land on which the majority of this work was completed. We thank the reviewer for their careful reading and constructive comments.

N.S. thanks Anshu Gupta and Arash Bahramian for useful discussions. C.L.H., I.H.W. and M.J. acknowledge support from the Oxford Hintze Centre for Astrophysical Surveys which is funded through generous support from the Hintze Family Charitable Foundation. C.L.H also acknowledges support from the Science and Technology Facilities Council (STFC) through grant ST/Y000951/1. R.A.M. acknowledges support from the Swiss National  Foundation (SNSF) through project grant 200020\_207349. I.S. acknowledges funding from the European Research Council (ERC) DistantDust (Grant No.101117541) and the Atracc\'{i}on de Talento Grant No.2022-T1/TIC-20472 of the Comunidad de Madrid, Spain. This research was supported by an Australian Government Research Training Program (RTP) Scholarship doi.org/10.82133/C42F-K220.  RB acknowledges support from an STFC Ernest Rutherford Fellowship [grant number ST/T003596/1].

The MeerKAT telescope is operated by the South African Radio Astronomy Observatory, which is a facility of the National Research Foundation, an agency of the Department of Science and Innovation. We acknowledge the use of the ilifu cloud computing facility – www.ilifu.ac.za, a partnership between the University of Cape Town, the University of the Western Cape, Stellenbosch University, Sol Plaatje University and the Cape Peninsula University of Technology. The Ilifu facility is supported by contributions from the Inter-University Institute for Data Intensive Astronomy (IDIA – a partnership between the University of Cape Town, the University of Pretoria and the University of the Western Cape, the Computational Biology division at UCT and the Data Intensive Research Initiative of South Africa (DIRISA). The authors acknowledge the Centre for High Performance Computing (CHPC), South Africa, for providing computational resources to this research project. 

Based on data products from observations made with ESO Telescopes at the La Silla Paranal Observatory under ESO programme ID 179.A-2006 and on data products produced by the Cambridge Astronomy Survey Unit on behalf of the VIDEO consortium.

The Hyper Suprime-Cam (HSC) collaboration includes the astronomical communities of Japan and Taiwan, and Princeton University. The HSC instrumentation and software were developed by the National Astronomical Observatory of Japan (NAOJ), the Kavli Institute for the Physics and Mathematics of the Universe (Kavli IPMU), the University of Tokyo, the High Energy Accelerator Research Organization (KEK), the Academia Sinica Institute for Astronomy and Astrophysics in Taiwan (ASIAA), and Princeton University. Funding was contributed by the FIRST program from the Japanese Cabinet Office, the Ministry of Education, Culture, Sports, Science and Technology (MEXT), the Japan Society for the Promotion of Science (JSPS), Japan Science and Technology Agency (JST), the Toray Science Foundation, NAOJ, Kavli IPMU, KEK, ASIAA, and Princeton University.

This paper makes use of software developed for Vera C. Rubin Observatory. We thank the Rubin Observatory for making their code available as free software at http://pipelines.lsst.io/.

This paper is based on data collected at the Subaru Telescope and retrieved from the HSC data archive system, which is operated by the Subaru Telescope and Astronomy Data Center (ADC) at NAOJ. Data analysis was in part carried out with the cooperation of Center for Computational Astrophysics (CfCA), NAOJ. We are honored and grateful for the opportunity of observing the Universe from Maunakea, which has the cultural, historical and natural significance in Hawaii.
\end{acknowledgements}

\vspace{5mm}
\facilities{JWST(NIRSpec), MeerKAT}

\software{astropy} \citep{astropy:2013,astropy:2018} 

\bibliography{bigbiblio}{}
\bibliographystyle{aasjournal}

\end{document}